\documentclass[aps,prl,twocolumn,superscriptaddress,linenumbers]{revtex4}
\usepackage{graphicx}
\usepackage{float}
\begin{document}
\title {Glassy clusters: Relations between their dynamics and characteristic features of their energy landscape}

\author{Sandip De}
\altaffiliation{IBM Semiconductor Research and Development Center, Bangalore-560045, India}
\affiliation{Department of Physics, Universit\"{a}t Basel, Klingelbergstr. 82, 4056 Basel, Switzerland}
\author{Bastian Schaefer}
\affiliation{Department of Physics, Universit\"{a}t Basel, Klingelbergstr. 82, 4056 Basel, Switzerland}
\author{Ali Sadeghi}
\affiliation{Department of Physics, Universit\"{a}t Basel, Klingelbergstr. 82, 4056 Basel, Switzerland}
\author{Michael Sicher}
\affiliation{Department of Physics, Universit\"{a}t Basel, Klingelbergstr. 82, 4056 Basel, Switzerland}
\author{D. G. Kanhere}
\affiliation{
Centre for Simulations and Modeling, University of Pune, 411007 India}
\author{Stefan Goedecker}
\affiliation{Department of Physics, Universit\"{a}t Basel, Klingelbergstr. 82, 4056 Basel, Switzerland}
\date{\today}


\begin{abstract}
Based on a recently introduced metric for measuring distances between configurations, 
we introduce distance-energy (DE) plots to characterize the potential energy surface (PES) of
clusters. Producing such plots is computationally feasible on the density functional (DFT) 
level since it requires only a set of a few hundred stable low energy configurations including the 
global minimum. By comparison with standard criteria based on disconnectivity graphs and on the dynamics 
of Lennard-Jones clusters we show that the DE plots convey the necessary information about the character 
of the potential energy surface and allow to distinguish between glassy and non-glassy systems. We 
then apply this analysis to real systems on the DFT level and show that both glassy and non-glassy 
clusters can be found in simulations. It however turns out that among our investigated clusters 
only those can be  synthesized experimentally which exhibit a non-glassy landscape. 
\end{abstract}

\maketitle

The features of the potential energy surface (PES)~\cite{walesbook} and the resulting  consequences for the  physical properties of a system 
are subject to intensive research. Because of the technological importance of glassy bulk materials, extended glassy 
systems have been studied extensively~\cite{{ediger},{cuglia},{biroli}}. 
During the last decades or so a number of advances have been made in understanding the nature of the glass transition using 
powerful simulation and analytical methods~\cite{{sas-still},{kob},{stillengerweber}}. However a number of issues such as non exponential relaxation processes, rapid growth of relaxation times with decreasing temperatures, the role of potential energy surface (PES) and 
configurational entropy and spatial heterogeneity  continue to be debated~\cite{{karmar},{sun}}. The qualitative understanding is based 
on the nature of the energy landscape~\cite{walesbook}. It was shown~\cite{middleton} that glassy systems 
have a large number of local minima of similar energy which are separated by barriers of various heights.  




Turning to finite systems, the electronic structure, equilibrium geometries and many properties of atomic cluster have also been studied extensively at various
levels of theory.
 The PES and related properties of the Lennard Jones (LJ) 
clusters with up to 1000 atoms are well understood~\cite{walesbook}. 
Atomic clusters are known to display size sensitive properties. 
For example, some clusters such as the $LJ_{55}$ (Lennard Jones cluster of 55 atoms of same type)  are structure seekers that exhibit a strong 
tendency to fall into their unique ground state~\cite{LJ55}, whereas others such as $LJ_{75}$ have a multi-funnel character 
which makes it much harder to fall into the ground state~\cite{LJ55}. 
By ground state we denote in this article the geometrical configuration corresponding to the global minimum of the PES.
For gold clusters the basic structural motif of the ground state can for instance change by the 
addition of a single atom~\cite{bao}. 
Ground state geometries frequently exhibit amorphous structures.~\cite{{na-melt1},{na-melt2},{na-ghazi}}. This can lead to a flat 
heat capacity in gallium and aluminum clusters, whereas highly symmetric clusters of the same material give 
a peaked heat capacity~\cite{{jarrold1},{ga-melt}}.

Though it is believed that a glassy landscape would also lead to glassy dynamics in clusters, 
the reported work has been rather sporadic and evidence in terms of dynamical behavior at low temperature is missing \cite{berry,nayak-berry,banerjee}.
One of the early attempts to seek glassy behavior in clusters  was by Rose and Berry in their study of $(KCl)_{32}$
clusters~\cite{berry}, and by Nayak, Jena and Berry~\cite{nayak-berry}.
In a more recent work, Banerjee and Dasgupta have investigated the dynamics of glass forming liquids using a master equation approach 
within a network model~\cite{banerjee}. Unfortunately their cluster was a structure seeker with a well defined ordered structure. 
Nevertheless they did  obtain clear  indications of glassy behavior by removing the low energy part of the spectrum. 
The standard approach to probe the glassy nature is via very long molecular 
dynamics (MD) runs at various temperatures. Although feasible for LJ clusters, this is prohibitively expensive for a realistic 
treatment using Density functional theory (DFT).
 An alternative is to characterize the PES  using the associated 
disconnectivity graphs~\cite{karplus} which shows the relation between the energy differences of the local minima
and the barrier heights. However determination of a large number of saddle points is computationally also very expensive at 
the DFT level.  For this reason studies on glassy clusters 
based on a realistic description, such as DFT, are virtually nonexistent. 

The present work has two main objectives. First we introduce a  novel approach based on  distance-energy (DE) analysis~\cite{ali}, 
and show that a DE plot represents the essential characteristics of a potential energy landscape. To establish this, we carry out long time MD and compute 
relevant dynamical susceptibility for two model LJ clusters. Second we demonstrate the utility of the approach by applying it 
to four clusters on the DFT level
and show that one cluster has a glassy character whereas the other ones are structure seekers.

\begin{figure}
	\includegraphics[width=0.35\textwidth]{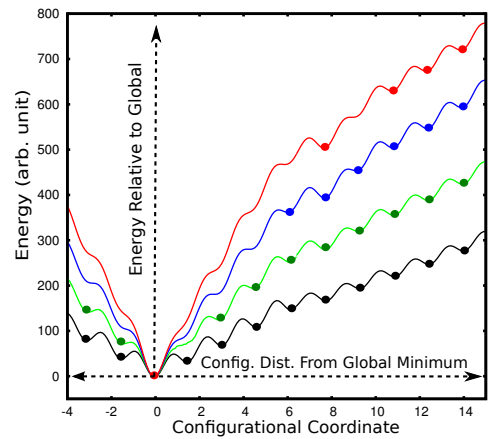}
\caption{\label{1dmodel} Simple one-dimensional model for the transformation of a glassy into a non-glassy energy landscape. 
The movement of the local minima, indicated by the discs, show the evolution of the DE plots during this transformation. }
\end{figure}

The basic idea is illustrated for a one-dimensional model in Fig~\ref{1dmodel}, where a 
glassy landscape is transformed into the landscape of a structure seeker by lowering the energy 
region around the global minimum with respect to the regions further away. During this transformation the energy 
differences between the global minimum and the  low energy local minima are obviously increased and 
some barriers disappear which in turn causes 
some local minima to disappear as well. This can  be explained mathematically 
by the Tomlinson model~\cite{tomlinson}. The DE plots for the PES at the four stages of the transformation are 
given by the locations of the local minima and shown by discs in the same color as the corresponding PES. 
Obviously the distance of a disk along the x axis is the distance of this local minimum from the global minimum 
in configurational space 
whereas the distance along the y axis is the energy of the local minimum with respect to the global minimum. 
For the structure seeker (red PES) the energy increases more rapidly with distance
and has fewer points which are close according 
to the configurational distance compared to the case of the glassy landscape (black discs). 

For realistic PESs which are very high-dimensional a suitable generalization 
of the distance is needed. 
A global fingerprint describing a cluster can be obtained from 
the eigenvalues of an overlap matrix of atom centered gaussians whose width is given by the 
covalent radius of the atom on which it is centered. 
The root mean square of the difference vector between two fingerprint vectors is then a distance measure which fulfills 
all the properties of a metric~\cite{ali}.
As we shall see it is this distance between the ground state and all metastable states along with their energies which reveals 
the character of a PES.
 Since for the LJ model systems the bond-length can not be 
approximated by the sum of the covalent radii, we use the following slightly modified matrix $C$ for the calculation of the eigenvalues of the LJ systems:
$C_{i,j} = \exp( -r_{i,j}^2/(2 \sigma_{ij}^2)) $, where $r_{i,j}$ is the distance between atom $i$ and $j$ and $\sigma_{ij}$ the 
parameter of the LJ potential (specified in the supplementary material) which takes on 3 different values depending on 
whether the atoms $i,j$ are of A or B type.
Since all the matrix elements used for the calculation of the configurational distance are scaled 
with respect to the  equilibrium bond-lengths, the configurational distance is independent of the bond-length 
and systems with different bond-lengths can be compared. Our results are rather insensitive to the exact functional form 
chosen for the calculation of the matrix elements $C_{i,j}$ and it is to be expected that even distances based on other descriptors 
of the chemical environments~\cite{bartok} will lead to similar results. 

The high dimensional character of the true PES leads to an important modification of the simple picture shown in Fig.~\ref{1dmodel}. 
Because local minima can be found in so many directions 
around the global minimum 
the number of minima within a certain configurational distance will be much larger than in our one-dimensional model 
and the density of points in the plot will be much higher. An even larger increase occurs for the saddle points which lead to 
neighboring minima. The global minimum of the $LJ_{13}$ cluster for instance is surrounded by 535 local minima 
which are connected to the global minimum by 911 saddle points~\cite{lj13}. 

Furthermore, there are 911 structurally
distinct transition states connecting 535 

We will next show that DE plots convey all the necessary information to judge whether a system has glassy character or not.
To do so we study two binary LJ systems (BLJ) having 45 (13 of type A and 32 of type B) and 55 (13 of type A and 42 of type B) 
atoms. The LJ potential parameters used for these two cluster are given in the supplementary material. 
Then we establish the glassy nature of the 55 atom cluster using standard tools such as disconnectivity graphs and 
dynamical susceptibilities obtained from 
molecular dynamics. The same examination of the 45 atom cluster on the other hand shows that it is a structure seeker. 
We next compute and examine the DE plots and will see that they give information which is in agreement with 
the information obtained by the previous methods. 

\begin{figure}[!h]
\includegraphics[width=0.4\textwidth]{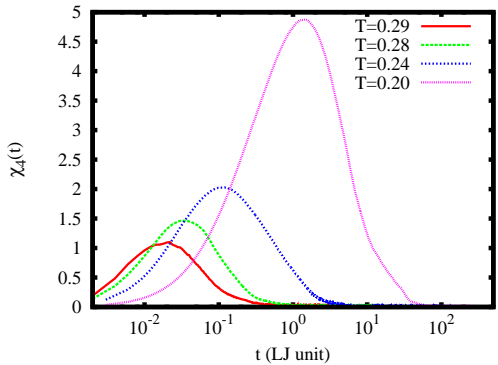}
\caption{\label{blj55x} Dependence of the dynamic susceptibility $\chi_4$ on time for different temperatures in the case of $BLJ_{55}$ 
}
\end{figure}

In order to compute long time dynamical properties
we have performed constant temperature MD using DLPOLY~\cite{dlpoly}
at five temperatures in the range $T \in [0.20,0.31]$.
The dynamics was studied via a two point correlation function~\cite{karmar}, $Q(t)$,
$$Q(t)=\int d\vec{r}\rho(\vec{r},t_0)\rho(\vec{r},t+t_0) \sim \sum_{i=1}^N w(|\vec{r}_i(t_0)-\vec{r}_i(t+t_0)|) \:, $$
where $\rho(\vec{r},t_0)$ are space-time dependent particle densities. $w(r)=1$, if $r \leq a$ and zero otherwise. 
The averaging over the initial time $t_0$ is implied. The window function $w$ of width $[a=0.30]$ treats 
particle positions separated by an amplitude smaller than .3 as identical. 
The dynamical susceptibility is defined as the fluctuation in $Q(t)$,
$\chi_4(t)=\frac{1}{N}[\langle Q^2(t) \rangle -\langle Q(t) \rangle^2]$.
It is well established that for glassy systems, $\chi_4(t)$ has a non-monotonic time dependence, 
and peaks at a time $\tau_4$ that is proportional to the
structural relaxation time. The time dependence of $\chi_4(t)$ is shown in Fig.~\ref{blj55x}  for $BLJ_{55}$. 
The note worthy feature is the increase in $\tau_m$ by two orders of magnitude as the temperature decreases. The behavior is 
very similar to the behavior of a glassy extended system ~\cite{karmar} and quantitatively establishes in the 
glassy character of the $BLJ_{55}$ cluster. 
On the other hand applying the same analysis to the $BLJ_{45}$ cluster 
does not give such a temperature dependence (see supplementary material).  

\begin{figure}[!h]
\includegraphics[width=0.35\textwidth]{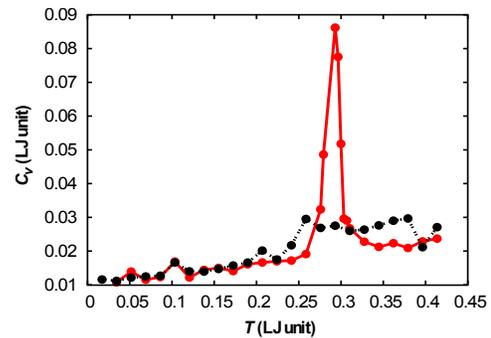}
\caption{\label{cv} The heat capacity $C_V$ as a function of temperature for the glassy cluster $BLJ_{55}$ (black curve) 
and the non-glassy $LJ_{55}$ one (red curve). }
\end{figure}

We also looked at an experimentally measurable fingerprint of glassy systems namely the heat capacity. 
As shown in Fig.~\ref{cv}, it is  rather flat for $BLJ_{55}$ and  shows no well 
defined peak, indicating the absence of a first order like transition. 
For comparison Fig.~\ref{cv} also shows the specific heat for another cluster of the same size, namely $LJ_{55}$, 
which is known to be a strong structure seeker. 
These calculated heat capacities are quite similar to the experimentally observed specific heats of 
$Ga_{30}$ and $Ga_{31}$ cations ~\cite{jarrold1}, which were termed as melters and non melters. The phenomenon has been explained 
on the basis of their respective geometries, the magic melters being relatively more ordered and non melters being disordered~\cite{ga-melt}. 
The same explanation applies in this case. The non-glassy LJ cluster is a icosahedral structure, whereas the glassy system is disordered. 

\begin{figure}[!h]
\includegraphics[width=\columnwidth]{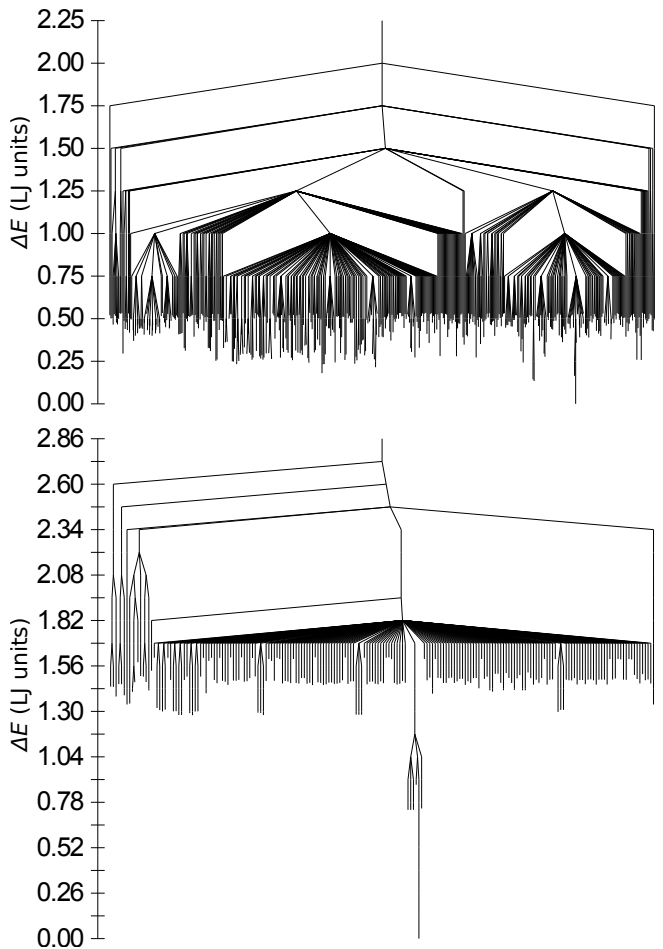}
\caption{ \label{disconnect} Disconnectivity graphs for the glassy BLJ55 (top) and non-glassy BLJ45 (bottom) binary Lennard Jones clusters.
The graphs were produced with the disconnectionDPS software \cite{DPS}. }
\end{figure}

Fig.~\ref{disconnect} shows the disconnectivity trees for our two model binary LJ clusters. The differences are obvious.
The structure seeker has a ground state which is considerably lower in energy than the next metastable configurations.
In the case of the glassy system there are many metastable configurations which are close in energy to the 
ground state and the barriers that have to be surmounted to get from one metastable configuration into another one are 
of variable height and frequently much larger than the energy difference between the local minima. 

\begin{figure}[!h]
	\includegraphics[width=0.35\textwidth]{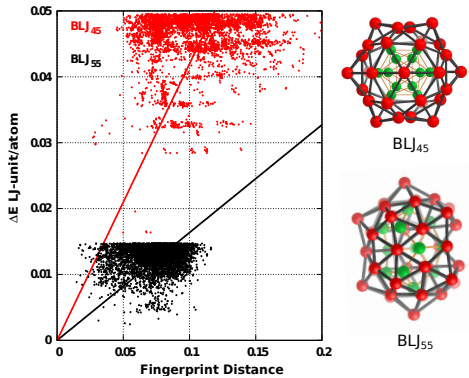}
\caption{\label{DELJ} 
DE plots for the glassy (black dots) and non-glassy (red dots) binary investigated Lennard Jones systems together with their ground state structures. Only the lowest 
5000 configurations are considered.
The two solid lines show least-square fits to the two data sets. Their slope is a measure for the average driving force towards the ground state. }
\end{figure}

Now we present and  discuss the DE plots (Fig.~\ref{DELJ}) for both systems.
The differences are striking and by comparison with our 
model PES of Fig~\ref{1dmodel}, it is clear that the $BLJ_{55}$ has a glassy PES whereas $BLJ_{45}$ does not have a glassy character. 
As has already be seen from the disconnectivity plot the global minimum is much lower in energy than any  other metastable state 
for the structure seeker. In addition the first metastable structure has also a rather large configurational distance from the global minimum. 
For the glassy system, on the other hand, there exists a large number of local minima close to the global minimum
which implies that the density of structures in the configurational space is higher for the glassy system.
Indirectly the large number of close-by minima also indicates that low saddle points exist around the global minimum. 
This is related to the Bell-Evans-Polanyi principle~\cite{bep} which states that barriers are low if the educt and product of a chemical reaction are similar. Hence the system has a 
distribution of low and high barriers characteristic of glassy systems.

\begin{figure}[!h]
	\includegraphics[width=0.35\textwidth]{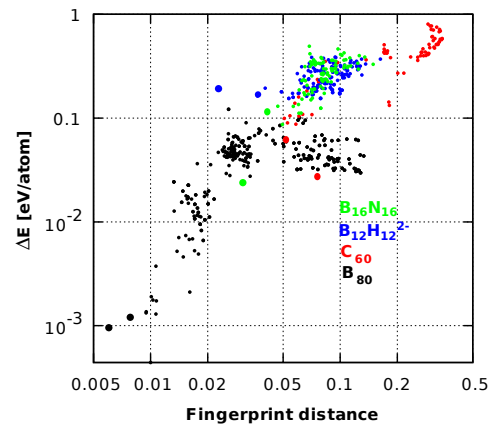}
	\caption{\label{DEdft} DE plots for a $B_{80}$, $C_{60}$ ,$B_{12}H_{12}^{2-}$ and $B_{16}N_{16}$ cluster. 
	Configurations close to the ground state are shown by larger discs.}
\end{figure}

We next demonstrate the capability of the method by investigating real clusters treated at the DFT level. 
We consider four clusters: $C_{60}$, $B_{12}H_{12}^{2-}$, $B_{16}N_{16}$ and $B_{80}$.
All the low energy metastable configurations required for the DE plots were found 
using the minima hopping method~\cite{minhop} coupled to the BigDFT electronic structure code~\cite{bigdft,PSP}.  
The DE plots are shown in Fig~\ref{DEdft}. Since the atomization energies per atom for all these 
covalently bonded system are of comparable magnitude (3 - 4 eV) no scaling of the energies was performed.  
The DE plots therefore show the energy per atom relative to the global minimum for all systems. 

We will first consider the $C_{60}$ cluster, whose experimental synthesis in 1985~\cite{c60} was considered 
a major breakthrough in chemistry. It is well known that the first local minimum corresponds to a 
Stone-Wales defect~\cite{stone} and that it is significantly higher in energy than the ground state, namely 
by about 1.6 eV. A disconnectivity graph based on DFT energies for saddle points that were found by 
a tight binding scheme has also been constructed for this system~\cite{walesc60} and 
it was found to be of the willow type, indicating that it is a structure seeker at higher temperatures. This explains the 
high temperature necessary for its experimental synthesis. In the DE plot the relatively high barriers can be deduced from 
the relatively large configurational distance of the lowest metastable structures from the global minimum. 
The second system is the chemically highly  stable icosahedral dodecahydro-closo-dodecaborate dianion $B_{12}H_{12}^{2-}$, 
which was already synthesized around 1960~\cite{b12h12}. 
Its structure seeker character can be deduced from the fact that the first metastable 
structure is much higher in energy than the ground state but not too far in configurational distance. This suggests that 
the barrier for jumping from the first metastable structure into the ground state is relatively small and that there is  
in general a strong driving force towards the ground state. 
The $B_{16}N_{16}$ has also been observed experimentally ~\cite{B16N16} and it also exhibits structure seeker features 
in its DE plot. 
A system which exhibits  a behavior quite opposite to the other ones is the $B_{80}$ cluster for which 
the lowest structure consists of a 12 atom icosahedron embedded in a disordered half dome structure~\cite{sandipB80}. This is the only system for which 
no well characterized experimental structure has been found.  Its DE plot is totally different form the other system. 
There are numerous metastable configurations which are both close in energy and close in configurational distance to the ground state. 
This means that there are many close-by configurations of similar energy separated by barriers of different heights. 
Thus we predict $B_{80}$ to be glassy.

In conclusion, we propose DE plots as a computationally tractable method to characterize a PES 
on the DFT level.They clearly show the strength of the driving force towards the ground state and 
they contain information about the density of metastable configurations in the configurational space. 
By applying it to several realistic clusters treated on the DFT level, we find clusters with both 
glassy and non-glassy behavior. All the investigated clusters that were synthesized experimentally exhibit a non-glassy 
energy landscape. This suggests that a landscape of this type is a prerequisite for experimental synthesis.



We thank the Indo-Swiss Research grant and SNF for the financial support and CSCS for computing time.
DGK acknowledges a number of discussions with Chandan Dasgupta.


\begin{thebibliography}{99}

\bibitem{walesbook}
D. J. Wales,Energy Landscapes,Cambridge University Press,
Cambridge, (2003).
\bibitem{ediger}
M. D. Ediger, Annu. Rev.Phys.Chem. 51, 99(2000)
\bibitem{cuglia}
Leticia F. Cugliandono, Les Houches Lecturer notes, arXiv: Cond Matter/0210312, ( 2002)
\bibitem{biroli}
G.Biroli,J.P Bouchard, A.Cavagna, T.S.Grigera, and P.Verrocchio, Nature Physics 4,771(2008)
\bibitem{sas-still}
Srikanth Sastri, Pablo G. Debendetti and Frank H. Stillinger, Nature 393,554(1998)
\bibitem{kob}
W.Kob and H.C. Andersen, Phy.Rev. E 51,4626(1995)
\bibitem {stillengerweber}
F.H. Stillinger and T.A. Weber, Phys. Rev. A 25, 978(1982),
F.H. Stillinger and T.A. Weber, Phys. Rev. A 28, 2408 (1983),
F.H.Stillinger, Science 267, 1935(1995)
\bibitem{karmar}
Smarajit Karmarkar, Chandan Dasgupta and Srikanth Sastry, Proc.Nat.Acad.Sc. 186, 3675(2009)
\bibitem{sun}
J. Sun, D.J. Earl and M.W. Deem, Phys. Rev. Lett. 95, 148104 (2005)
\bibitem{middleton}
	T.F. Middleton and D. Wales, Phys. Rev. B 118 4583 (2003)
\bibitem{LJ55}
Jonathan P. K. Doye, Mark A. Miller, and David J. Wales J. Chem. Phys. 111, 8417 (1999)
\bibitem{bao}
Kuo Bao, Stefan Goedecker and Kenji Koga, et al., Phys. Rev. B 79, 41405(2009)
\bibitem{na-melt1}
S. Chacko, D.G. Kanhere, and S.A. Blundell, Phys.Rev. B 71, (15), 155407(2005)
\bibitem{na-melt2}
M.S. Lee, S.Chacko, and D.G. Kanhere, J.Chem. Phys.123, 1643105(2006)
\bibitem{na-ghazi}
S.M.Ghazi, S.Zorriasatein, and D.G.Kanhere, J.Phys.Chem. A 113, 2659(2009)
\bibitem {jarrold1}
Gary A. Breaux, Damon A. Hillman, Colleen M. Neal, Robert C. Benirschke, and Martin F. Jarrold,
J. Am. Chem. Soc.126, 8629(2004)
\bibitem{ga-melt}
K.Joshi, S.Krishnamurty, and D.G. Kanhere, Phys. Rev. lett. 96, 135703(2006)
\bibitem{berry}
John P. Rose and R. Stephen Berry, J. Chem. Phys. 98, 3262 (1993)
\bibitem{nayak-berry}
Saroj K. Nayak, Puru Jena,Keith D. Ball and R.Stephen Berry, J. Chem. Phys. 108 , 234 (1998)
\bibitem {banerjee}
Sumilan Banerjee and Chandan Dasgupta, Phys.Rev. E 85, 021501 (2012)
\bibitem{karplus}
O. M. Becker and M. Karplus, J. Chem. Phys. 106, 1495 (1997)
\bibitem{ali}
Ali Sadeghi, S Alireza Ghasemi, Bastian Schaefer, Stephan Mohr, Markus A Lill, Stefan Goedecker, J. Chem. Phys. 139, 184118 (2013)
\bibitem{tomlinson}
G.A. Tomlinson, Phil. Mag, 7, 905(1929)
\bibitem{bartok}
Bartok, Albert P., Kondor, Risi, Csanyi, Gabor, Phys. Rev. B 87, 219902 (2013) 
\bibitem{lj13}
	J. P. K. Doye, M. A. Miller and D. J. Wales, J. Chem. Phys. 111, 8417 (1999)
\bibitem{dlpoly}
I.T. Todorov, W. Smith, K. Trachenko and M.T. Dove, Journal of Materials Chemistry, (2006) 16, 1911-1918
\bibitem{DPS}
M. Miller, D. Wales, and V. de Souza, 
http://www-wales.ch.cam.ac.uk/~wales/disconnectionDPS.f90
\bibitem{bep}
F. Jensen, {\it Computational Chemistry}, Wiley, New York (1999).
\bibitem{minhop}
Stefan Goedecker, J. Chem. Phys. 120, 9911 (2004)
\bibitem{bigdft}
Luigi Genovese, Alexey Neelov, Stefan Goedecker, Thierry Deutsch, Alireza Ghasemi,
Oded Zilberberg, Anders Bergman, Mark Rayson and Reinhold Schneider,
J. Chem. Phys. 129, 014109 (2008)
\bibitem{PSP}
S. Goedecker, M. Teter, J. Hutter, Phys. Rev. B {\bf 54}, 1703 (1996) ; 
A. Willand, Y.O. Kvashnin, L. Genovese, A. V\'azquez-Mayagoitia, A.K. Deb, A.Sadeghi, T, Deutsch and S. Goedecker,
J. Chem. Phys. {\bf 138}, 104109 (2013)

\bibitem{c60}
H.W. Kroto, J.R. Heath, S.C. O'Brien, R.F. Curl, R.E. Smalley, Nature 318,162(1985).

\bibitem{stone}
A. Stone and D. Wales, Chem. Phys. Lett. 128, 501(1986)

\bibitem{walesc60}
Y. Kumeda, and D.J. Wales, Chem. Phys. Lett. 374, 125(2003)

\bibitem{b12h12}
Anthony R. Pitochelli, Frederick M. Hawthorne, J. Am. Chem. Soc. 82, 3228(1960)

\bibitem{B16N16}
Zhi Xu, Dmitri Golberg, Yoshio Bando, Chem. Phys. Lett. 480, 110 (2009)

\bibitem{sandipB80}
Sandip De, Alexander Willand, Maximilian Amsler, Pascal Pochet, Luigi Genovese, and Stefan Goedecker,
Phys. Rev. Lett. 106, 225502 (2011)



\end{thebibliography}
\end{document}